\documentstyle[12pt,prb,aps]{revtex}
\tiny\normalsize 
\newcommand{\tilt}{\tilde{t}}
\newcommand{\tilx}{\tilde{x}}

\input epsf

\begin{document}
\draft

\title{Decay of one dimensional surface
modulations}
\author{Navot Israeli\cite{NavotEmail} and Daniel
Kandel\cite{DanielEmail}}
\address{Department of Physics of Complex Systems,\\
Weizmann Institute of Science, Rehovot 76100, Israel}
\maketitle

\begin{abstract}
The relaxation process of one dimensional surface modulations is
re-examined. Surface evolution is described in terms of a standard 
step flow model. Numerical evidence that the surface slope, $D(x,t)$,
obeys the scaling ansatz $D(x,t)=\alpha(t)F(x)$ is provided. We use
the scaling ansatz to transform the discrete  step model into a
continuum model for surface dynamics. The model consists of differential
equations for the functions $\alpha(t)$ and $F(x)$. The solutions of
these equations agree with simulation results of the discrete step 
model. We identify two types of possible scaling solutions. Solutions of
the first type have facets at the extremum points, while in solutions of
the second type the facets are replaced by cusps. Interactions between
steps of opposite signs determine whether a system is of the first or
second type. Finally, we relate our model to an actual experiment and
find good agreement between a measured AFM snapshot and a solution of
our continuum model.
\end{abstract}


\section{Introduction}  
Periodic crystalline surfaces are a convenient testing
ground for modeling surface evolution and considerable
efforts have been devoted to their study. The decay of one and two
dimensional surface modulations has been investigated extensively.
The emerging experimental picture is that above the
roughening transition  
temperature, $T_R$, such structures decay as pure sine
waves\cite{Yamashita}. This fact is well explained by Mullins' theory
\cite{Mullins} which treats the surface as an isotropic object.  
However, below $T_R$ the situation is different and one must take into
account the anisotropy of
the underlying crystal lattice. This anisotropy produces cusp
singularities in the crystal surface tension at high symmetry
planes, which lead to the formation of facets
\cite{Yamashita,BlakelyUmbachTanaka_snapshoot,Tanaka,Blakely,Umbach}.  

From the theoretical point of view, the existence of cusps in the
surface tension 
complicates the description of surface evolution
in terms of continuous models. Such models usually assume that
mass transfer follows gradients in the {\em continuous} surface
chemical potential. However, the chemical potential is singular at facet
edges 
and the dynamics in the vicinity of these regions must be
treated with special care. 

There are basically two ways to deal with
the above problem. The first  
one is to avoid it all together by resorting to microscopic models. Step
flow models
\cite{OzdemirZangwill,RettoriVillain,DuportChameMullinsVillain} and
Monte Carlo simulations
\cite{RamanaCooper,AdamChameLanconVillain,DubsonJeffers,JiangEbner,%
SelkeDuxbury,ErlebacherAziz} describe
surface dynamics in terms of {\em discrete} microscopic objects. On
this level the surface chemical potential is no longer continuous and
there are no singularity problems. Facets are
natural phenomena in such a framework. However in going to microscopic
models one usually loses the mathematical simplicity of the continuum
approach. Step flow models and Monte Carlo simulations are usually
solved numerically and it is difficult to derive the large scale
dynamics
from them. 

The second way is to use continuum models in which singular points are
treated as
moving \cite{HagerSpohn} or stationary
\cite{OzdemirZangwill} boundaries. Alternatively, surface
tension singularities can 
be smoothed out \cite{BonzelPreussSteffen,BonzelPreuss} assuming that if  
smoothing is done on a small length scale the large scale results will 
not be affected. The drawback of these continuum approaches is that by
starting with a continuum model one loses sight of the underlying
microscopic kinetics. As pointed out by Chame et
al.\cite{ChameRoussetBonzelVillain}, the connection of these models with
the details
of the microscopic processes is unclear. We show below that
these details are important and may lead to large scale effects. 

In this work we re-examine the relaxation process of unidirectional
surface modulations. Our aim is to derive a continuum model for the
surface evolution which is consistent with the microscopic
kinetics. We adopt here the same approach we used in the
study of other surface
structures\cite{cone_short,cone_long,1D_scaling}. First we study the 
system's behavior  
in terms
of a standard step flow 
model. The model allows steps of opposite 
sign to interact; such interactions can lead to facet
formation. Numerical simulations of the step model suggest that
the surface slope $D(x,t)$ follows the simple scaling law 
$D(x,t)=\alpha(t)F(x)$. The derivation of the step model and its
numerical 
analysis are carried out in Section II. In Section III we use the
scaling
ansatz to transform the 
step flow model into a continuum model, i.e., we derive the
differential equations for the functions  $\alpha(t)$ and
$F(x)$ directly from the step velocities of the discrete model. 
In Section IV we solve the
resulting differential equations and find the scaling function $F$. 
We find an impressive agreement between this scaling function and
simulations of the step model.

In Section V we discuss the relevance of our model to
AFM measurements of decaying surface modulations taken by Tanaka et
al.\cite{BlakelyUmbachTanaka_snapshoot}. We show that the best fit
solution of
our model agrees with the experimental data. On this basis we derive a
formula which connects microscopic parameters in the system with the
measured life time of the profile.
Our conclusions and their relation to existing work are presented in
section VI.

\section{Step flow model of 1D modulations}
\begin{figure}[h]
\centerline{
\epsfxsize=90mm
\epsffile{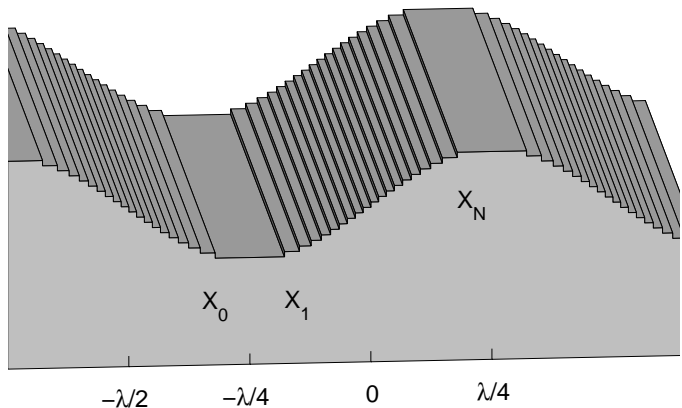}}  
\caption{Illustration of a modulated surface below the roughening
transition. The step labeling is indicated. We label the
terraces such that the $n$th terrace separates the $n$th and $n+1$th
steps.}  
\label{sine_picture}
\end{figure}

Consider a 1D grating of wavelength $\lambda$ fabricated on a high
symmetry crystal 
orientation as illustrated in Fig.\ \ref{sine_picture}. Below the   
roughening transition one can ignore the existence of islands and
voids and describe the  
surface as consisting of flat terraces separated by straight atomic
steps. In the absence of evaporation and when no new material is added,
these steps move by mass exchange with the adatom diffusion
fields on neighboring terraces. Such
evolution was treated 
in the spirit of the Burton-Cabrera-Frank model\cite{BCF} by a number of 
authors. In what follows we use a standard
step model for surface evolution. This model is essentially
equivalent to the one used by Ozdemir and Zangwill in
Ref. \onlinecite{OzdemirZangwill} except that we allow steps of opposite
sign
to interact. For completeness we now derive the model.

To describe step motion mathematically, one has to solve the adatom
diffusion problem on the terraces. Let us denote the 
adatom concentration field on the $n$th terrace by $C_n(x)$. In most
situations, the time scale associated with step motion is much larger
than the time scale of surface diffusion. One can therefore assume
that the adatom concentration field is always in a steady state; i.e,
for 
any given step configuration, $C_n(x)$ reaches a steady state before
the steps move significantly. Within this quasi-static approximation 
$C_n(x)$ obeys the static
diffusion equation $\nabla^2C_n(x)=0$, which is readily solved by
\begin{equation}
C_n(x)=a_n+b_nx\;.
\label{concentration}
\end{equation}

Next we look at the boundary conditions for these diffusion fields.
Near the step edges the diffusion currents are determined by the flux
of atoms emitted or absorbed by the steps.
Assuming linear kinetics characterized by an
attachment-detachment  
rate $k$, the current of atoms on both sides of the $n$th step is given
by\cite{BalesZangwill}  
\begin{eqnarray}
J_{n-1}&=&-D_s \nabla
C_{n-1}\left(x_n\right)=k\left[C_{n-1}\left(x_n\right)-C_n^{eq}\right]\;
, \nonumber \\
J_n&=&-D_s \nabla
C_n\left(x_n\right)=-k\left[C_n\left(x_n\right)-C_n^{eq}\right]\;.
\label{diffusion_boundaries}
\end{eqnarray}
Here $D_s$ is the adatom diffusion coefficient,  $x_n$ is the
position of the $n$th step, $J_n$ is the current 
across the $n$th terrace and $C_n^{eq}$ is the equilibrium adatom
concentration at $x_n$.

Eqs.\ (\ref{concentration}) and (\ref{diffusion_boundaries}) can be
used to calculate the constants $a_n$ and $b_n$ and thus fully
determine the adatom concentration fields. We find that 
\begin{equation}
J_n=-D_sb_n=\frac{D_s\left(C_n^{eq}-C_{n+1}^{eq}\right)}{\frac{2D_s}{k}+
x_{n+1}-x_n}\;.
\label{current}
\end{equation}  
Mass conservation implies that the velocity of the $n$th step takes
the form  
\begin{equation}
\frac{dx_n}{dt}=\Omega \left(J_n-J_{n-1}\right)\;,
\label{velocities1}
\end{equation}
where $\Omega$ is the atomic area of the solid.

In order to complete the model, we write down
expressions for $C_n^{eq}$. To this end we introduce the step chemical
potential 
$\mu_n$, which is associated with the addition of an atom to the solid
at the
$n$th step. $C_n^{eq}$ depends on the step chemical potential in the
following way: 
\begin{equation}
C_n^{eq}=\tilde{C}^{eq}\exp \frac{\mu_n}{k_B T} \approx
\tilde{C}^{eq}\left(1+\frac{\mu_n}{k_B T}\right)\;.
\label{equilibrium_concentration}
\end{equation}
$\tilde{C}^{eq}$ is the equilibrium concentration of a noninteracting
step, $k_B$ is the Boltzmann constant and $T$ is the temperature.

Finally, to evaluate $\mu_n$ we take into account repulsive
interactions with strength $\beta/2$ between nearest neighbor steps
of the same sign:  
\begin{equation}
U\left(x_n,x_{n+1}\right)=\frac{\beta}{2\left(x_{n+1}-x_n\right)^2}.
\label{repulsive_int}
\end{equation}
Such a repulsion is consistent with entropic as well as elastic
\cite{MarchenkoParshin,AndreevKosevich} interactions between straight
steps. We also take into account the possibility of an attractive
interaction with strength $\tilde{\beta}/2$ between nearest neighbor
steps of opposite signs: 
\begin{equation}
\tilde{U}\left(x_n,x_{n+1}\right)=-\frac{\tilde{\beta}}{2\left(x_{n+1}-x
_n\right)^2}.
\label{attractive_int}
\end{equation}
An elastic attraction of this form may exist in some
materials\cite{MarchenkoParshin}. In addition, step-antistep
annihilation events are accelerated by step
fluctuations\cite{DuportChameMullinsVillain,ChameRoussetBonzelVillain}.
In our one dimensional
model, step fluctuations are not treated. Instead, this kinetic
effect can be introduced as an effective step-antistep attraction.
Although
the form of this kinetic 
attraction may be different, Eq.\ (\ref{attractive_int}) can be viewed 
as a phenomenological ingredient of our model which enables us to
study the effect of step-antistep attraction on surface evolution.
Using the above interactions the chemical potentials of steps
$1,\dots,N$ in Fig.\ \ref{sine_picture} are given by 
\begin{eqnarray}
\mu_1&=&\frac{\partial
\left[\tilde{U}\left(x_0,x_1\right)+U\left(x_1,x_2\right)\right]}
{\partial x_1}=
\frac{\beta}{\left(x_2-x_1\right)^3}+\frac{\tilde{\beta}}{\left(x_1-x_0
\right)^3}\;, \nonumber \\
\mu_n&=&\frac{\partial
\left[U\left(x_{n-1},x_n\right)+U\left(x_n,x_{n+1}\right)\right]}
{\partial x_n}=
\frac{\beta}{\left(x_{n+1}-x_n\right)^3}-\frac{\beta}{\left(x_n-x_{n-1}
\right)^3},\;\;\;
n=2, \ldots, N-1, 
\nonumber \\
\mu_N&=&\frac{\partial
\left[U\left(x_{N-1},x_N\right)+\tilde{U}\left(x_N,x_{N+1}\right)\right]
}{\partial x_N}=
-\frac{\beta}{\left(x_N-x_{N-1}\right)^3}-\frac{\tilde{\beta}}{\left(x_{
N+1}-x_N\right)^3}\;.
\label{chemical_potentials}
\end{eqnarray}

Eqs.\ (\ref{current}), (\ref{velocities1}),
(\ref{equilibrium_concentration}) and  (\ref{chemical_potentials})
constitute a complete model for surface  
evolution. This model depends on six parameters but many of them are
trivial. The only non trivial parameters in
the model are the ratio $g=\tilde{\beta}/\beta$, which measures the
relative strength of the two interactions, and the length
$l=\frac{2D_s}{k}$.   
The latter determines the rate limiting
process in the system.  
When $l \rightarrow 0$ attachment-detachment events are relatively
fast and the kinetics is diffusion limited 
(DL). In the  
opposite case, $l \rightarrow \infty$, diffusion is relatively fast
and the kinetics is attachment-detachment  
limited (ADL). 
The combined effect of all the other parameters can be scaled out by
transforming to dimensionless variables. In the DL case we choose  
\begin{eqnarray}
\tilx&=&\frac{2\pi}{\lambda}x\;, \nonumber \\
\tilt&=&\left(\frac{\lambda}{2\pi}\right)^5 \cdot \frac{k_B T}{\Omega
D_s \tilde{C}^{eq} \beta}t\;.
\label{DL_dimensionless_vars}
\end{eqnarray}
In all other non-diffusion-limited (NDL) cases we choose
\begin{equation}
\tilt=\left(\frac{\lambda}{2\pi}\right)^4 \cdot \frac{k_B T}{\Omega k
\tilde{C}^{eq} \beta}t\;.
\label{NDL_dimensionless_vars}
\end{equation}
with the same definition of $\tilx$. In the remainder of this section we
study
surface evolution through numerical  
simulations of the above model in terms of the dimensionless variables
$\tilx$ and $\tilt$.

\begin{figure}[h]
\centerline{
\epsfxsize=90mm
\epsffile{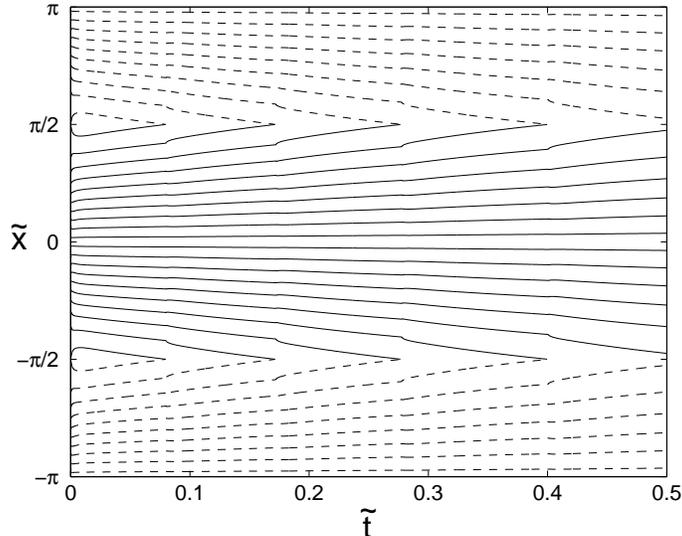}}  
\caption{Simulation result of a typical step system. Solid (dashed)
curves show the position of steps (antisteps). Steps and antisteps
collide and annihilate 
at the profile maximum ($\tilx=\pi/2$) and minimum ($\tilx=-\pi/2$).} 
\label{typical_simulation}
\end{figure}

Figure \ref{typical_simulation} presents the simulation results of a
typical system.  
We show a plot of the step configuration in one wave 
length of the profile. The initial step configuration corresponds to a
sinusoidal surface profile.
After a short transient step-antistep pairs at 
the top and bottom terraces start annihilating. As a result, the steps
in the  
sloping parts of the profile become less densed.

To further understand the surface evolution let us study step
kinetics through their density function (the profile slope). We  
define the step density between two neighboring steps of the same sign
as the 
inverse step separation, i.e.,
\begin{equation}
D\left(\frac{\tilx_{n+1}+\tilx_n}{2},\tilt\right)=\pm
\frac{1}{\tilx_{n+1}(\tilt)-\tilx_n(\tilt)}\;,
\label{density_def}
\end{equation}  
where the $\pm$ signs distinguish between steps and antisteps.
The density between two steps of opposite sign is defined to be zero,
consistently with the profile slope.  

$D(\tilx,\tilt)$ is a continuous function of $\tilt$, defined at a  
discrete set of positions at any given time. It turns out that the
evolution of this
function obeys a  
simple scaling ansatz. When we scale the step density to unity at its
peak and 
then plot density functions from different times, all the data collapses
onto a single  
curve. In other words, there exist functions $F(\tilx)$ and
$\alpha(\tilt)=\max_{\{\tilx\}} D(\tilx,\tilt)$, which satisfy the
relation 
\begin{equation}
D\left(\tilx,\tilt\right)=\alpha \left( \tilt \right) F \left( \tilx
\right)\;,
\label{scaling_ansatz}
\end{equation}
for every $\tilx=\left(\tilx_{n+1}+\tilx_n\right)/2$. Note that since
the step positions vary 
continuously with time, $F$ is a function of a {\em continuous}
variable.

\begin{figure}[h]
\centerline{
\epsfxsize=80mm
\epsffile{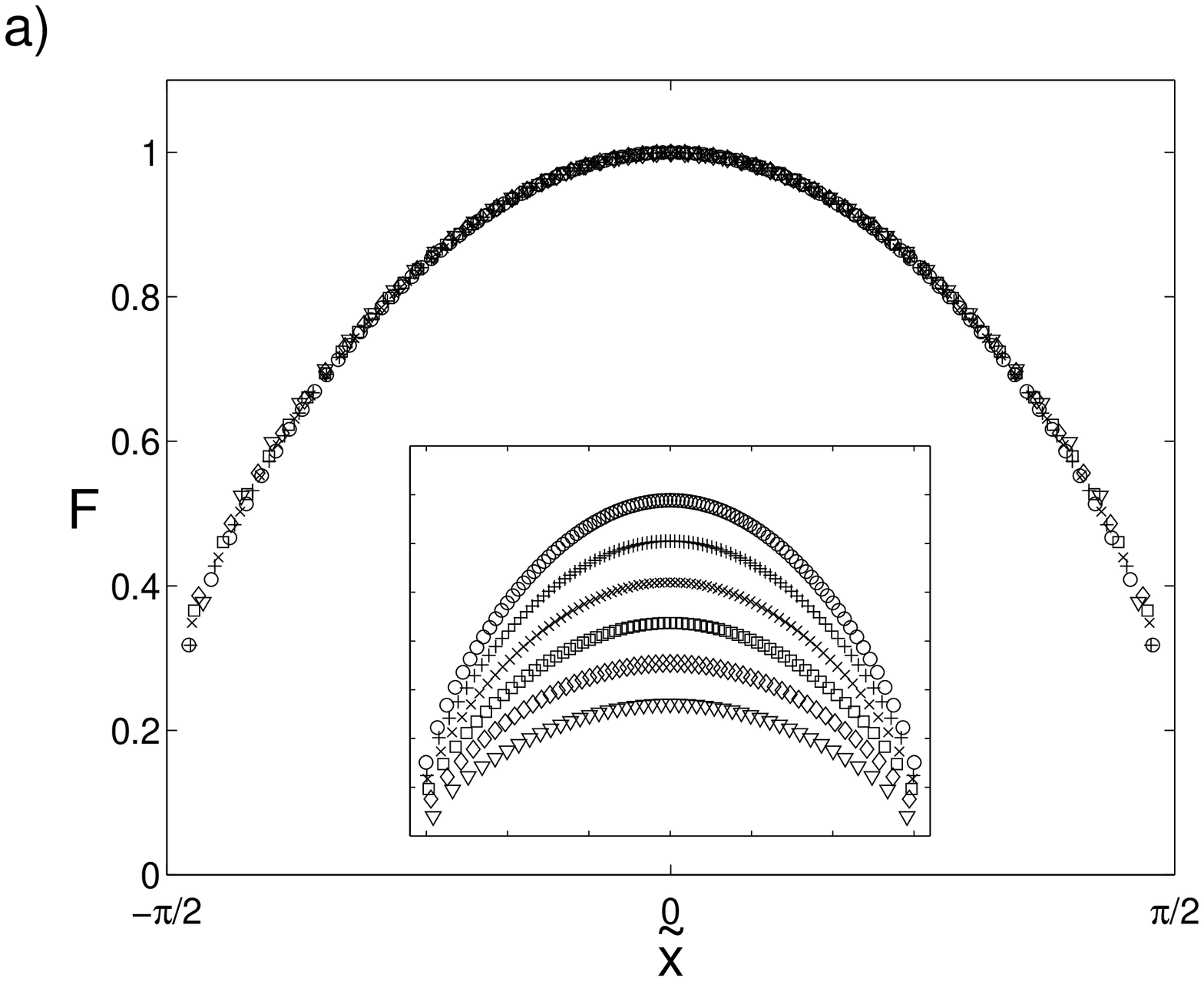}
\hspace{0.1in}
\epsfxsize=80mm
\epsffile{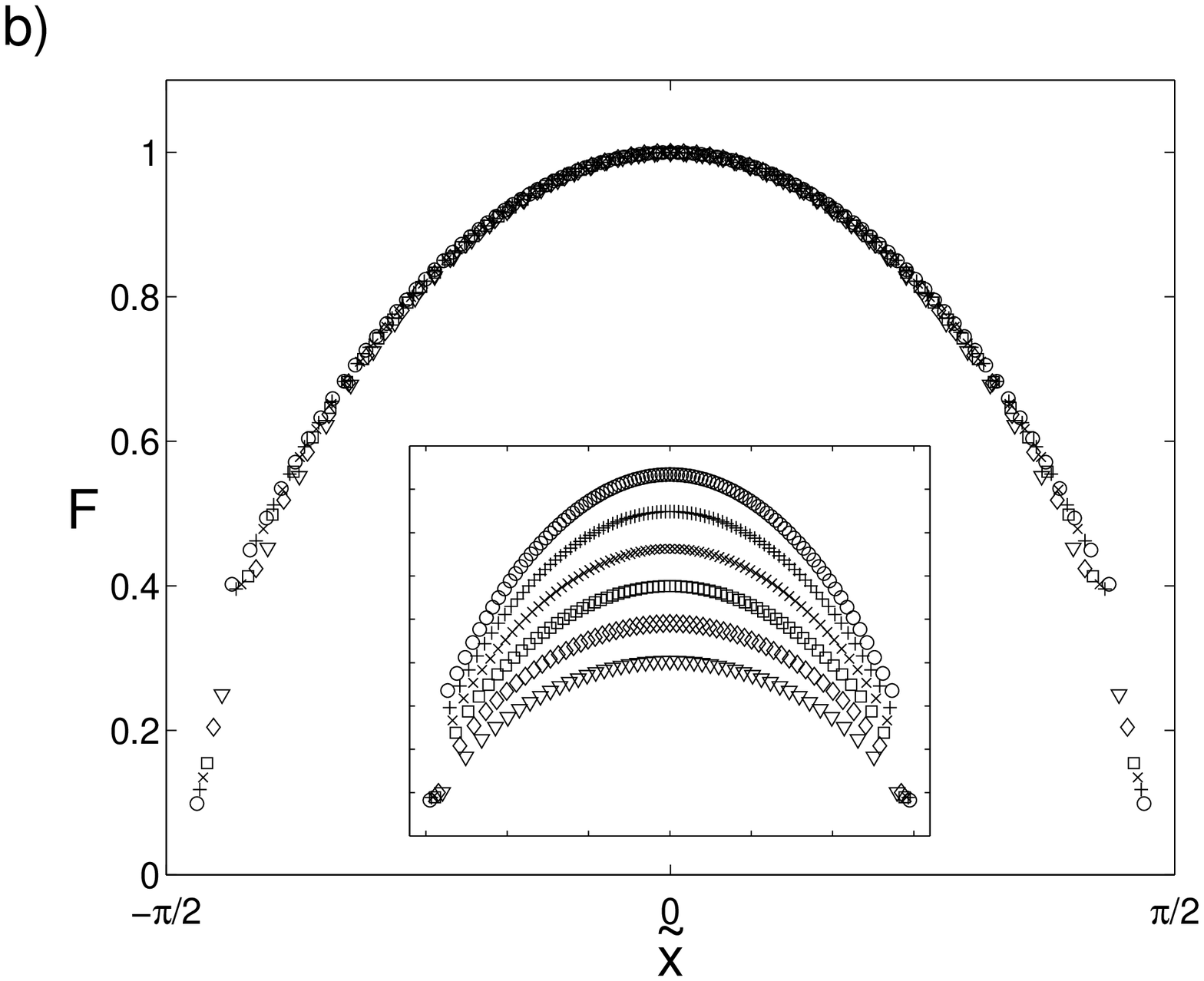}}
\caption{Data collapse of density functions in the ADL case with a)
$g=0$ and b) $g=24$. Note that in the
presence of step-antistep attraction ($g=24$) the density function is
steeper and has a smaller value near the profile extrema.}
\label{figure3}
\end{figure}

Figure \ref{figure3} demonstrates the data   
collapse obtained when the density function is divided by its
amplitude, $\alpha(\tilt)$, at different   
times for the ADL case with (Fig.\ \ref{figure3}a) and without (Fig.\
\ref{figure3}b) step-antistep attraction. Only 
half a period of a profile modulation is shown, since by 
symmetry the
density of antisteps in the other half behaves in exactly the same
way. A similar behavior is seen in the DL limit. We note that in all
cases the system exhibits scaling to a good approximation, but the
quality of the data collapse 
is slightly better  
in the cases without step-antistep attraction ($g=0$). The second
observation is that when
step-antistep attraction is present, the scaling function is steeper and
has a smaller step density near the profile extrema. This is a 
result of the faster step-antistep annihilation process due to the
attraction.

\begin{figure}[h]
\centerline{
\epsfxsize=80mm
\epsffile{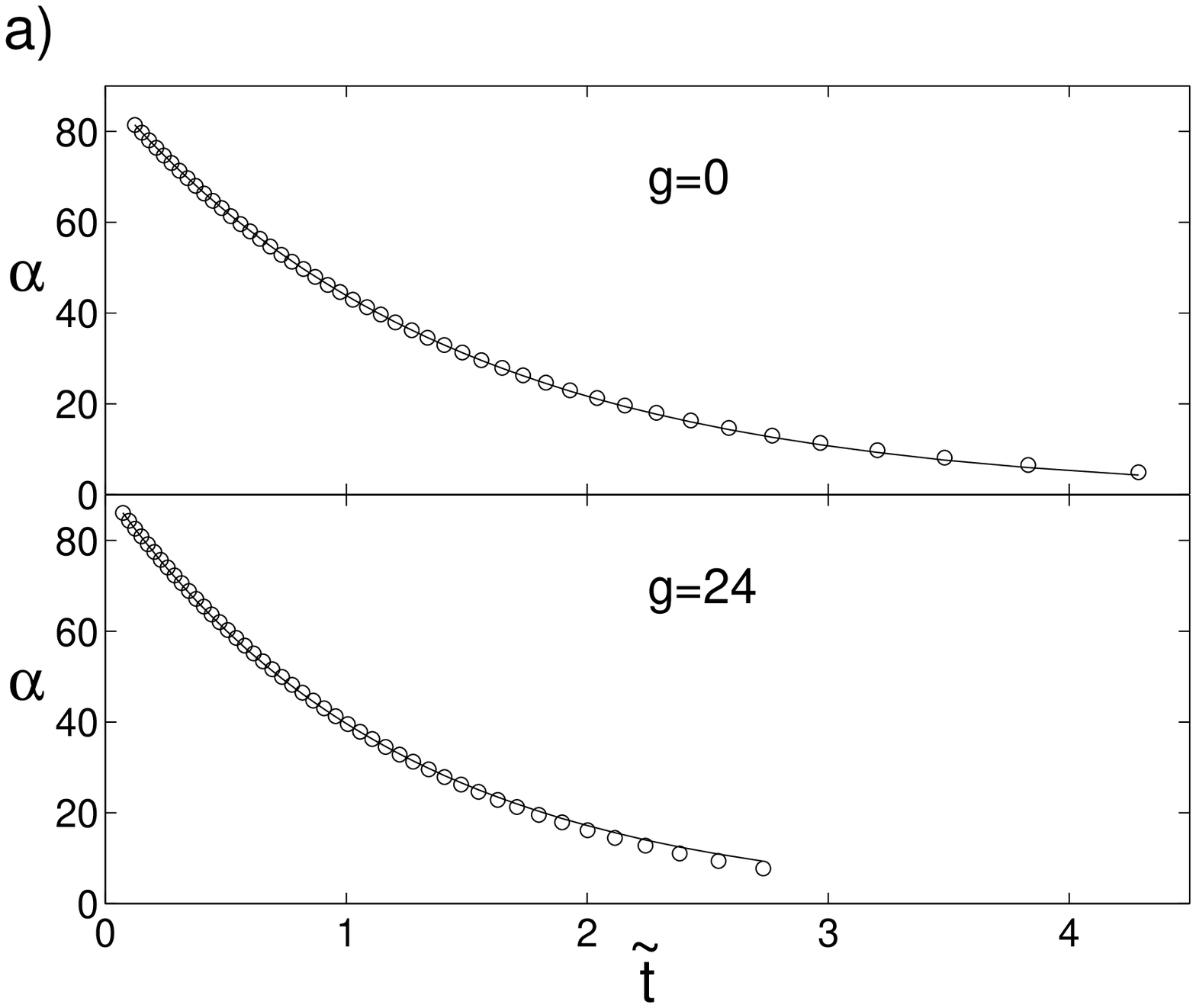}
\hspace{0.1in}
\epsfxsize=80mm
\epsffile{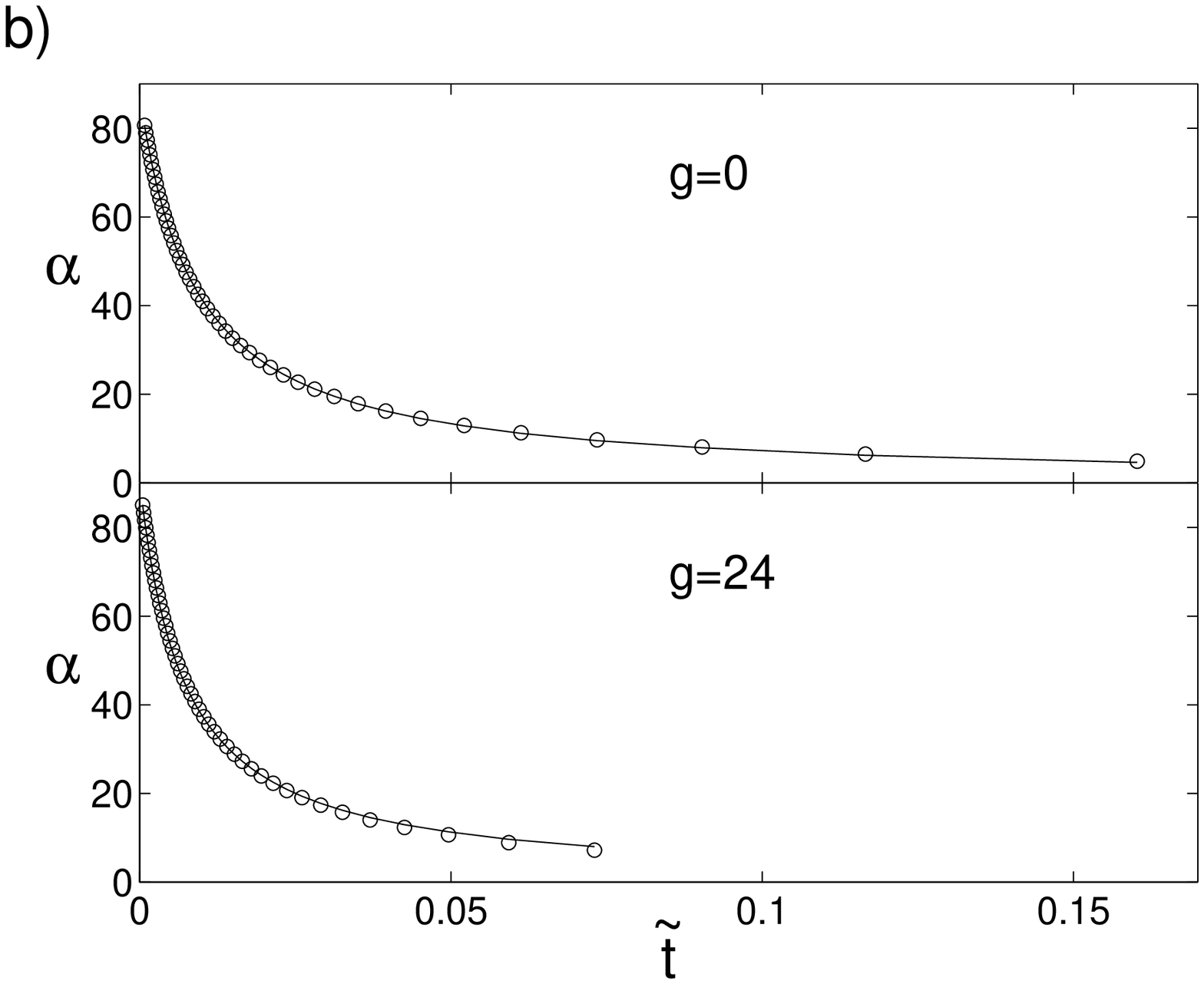}}
\vspace{0.1in}  
\caption{The amplitudes $\alpha(\tilt)$ in different systems
(circles): a) ADL with $g=0$ and $g=24$, b) DL   
with $g=0$ and $g=24$. In the ADL cases the 
amplitudes can be fitted (solid lines in a) by
$\alpha=A*e^{-\tilt/\tau}$ with 
$\tau=1.42$ for the $g=0$ case and $\tau=1.196$ for the $g=24$
case. In the DL cases the  
amplitudes can be fitted (solid lines in b) by
$\alpha=\tau/(\tilt-\tilt_0)$ with 
$\tau=0.785$ for the $g=0$ case and $\tau=0.643$ for the $g=24$ case.
Note that in the
presence of step-antistep attraction the decay is faster.}
\label{figure4}
\end{figure}

While there is no significant difference between the ADL and DL
scaling functions, the time dependent amplitude, $\alpha(\tilt)$, does
show
different behaviors. In Fig.\ \ref{figure4} 
we plot $\alpha(\tilt)$ for ADL and DL systems, with and without
step-antistep attraction. In the ADL cases, 
$\alpha(\tilt)$ can be fitted by an exponential decay, while in the DL
cases $\alpha(\tilt) 
\sim (\tilt-\tilt_0)^{-1}$ consistently with Ref.
\onlinecite{OzdemirZangwill}. 

\section{Scaling analysis and a continuum model}
The simulation results from the previous section suggest that the step
density in   
the discrete step model can be described by the continuous functions
$\alpha(\tilt)$  
and $F(\tilx)$. This situation is similar in many aspects to the one
encountered
in a different step  
system which describes the decay of a crystalline cone
\cite{cone_short,cone_long}. Although the cone  
system obeys a different scaling scenario, we use here the same
scaling analysis, described in detail in Ref. \onlinecite{cone_long}.
Our aim is to
derive the differential equations for the functions $\alpha(\tilt)$ and
$F(\tilx)$.  

Assuming that the scaling ansatz (\ref{scaling_ansatz}) indeed holds, we 
write the full time derivative of the step density:
\begin{equation}
\frac{dD}{d\tilt}=\frac{\partial D}{\partial \tilt}+\frac{\partial
D}{\partial 
\tilx}\cdot \frac{d\tilx}{d\tilt}\;.
\label{density_evolution1}
\end{equation}
Equation (\ref{density_evolution1}) can be evaluated in the middle of
the terrace 
between two steps, i.e. at $\tilx=(\tilx_n+\tilx_{n+1})/2$. The time 
derivative of $\tilx$ is then given by 
$d\tilx/d\tilt=(\dot{\tilx}_n+\dot{\tilx}_{n+1})/2$ where
$\dot{\tilx}_n\equiv d\tilx_n/d\tilt$. The l.h.s. of Eq. 
(\ref{density_evolution1}) can be calculated by taking the time
derivative of Eq. 
(\ref{density_def}):
$dD/d\tilt=-D^2\left( \dot{\tilx}_{n+1}-\dot{\tilx}_n \right)$. Using
the scaling ansatz,  
(\ref{scaling_ansatz}), we express Eq. (\ref{density_evolution1}) in
terms 
of the functions $\alpha(\tilt)$ and $F(\tilx)$:
\begin{equation}
\alpha F'\frac{\dot{\tilx}_{n+1}+\dot{\tilx}_n}{2} 
+\dot{\alpha}F+\alpha^2F^2\left(\dot{\tilx}_{n+1}-\dot{\tilx}_n\right)=0
\;.
\label{density_evolution2}
\end{equation}
$\dot{\alpha}$ and $F'$ are the $\tilt$ and $\tilx$ derivatives of
$\alpha$ and $F$, 
respectively.
The step velocities $\dot{\tilx}_n$ and $\dot{\tilx}_{n+1}$ can in
principle
be expressed in terms of the $\tilx_n$'s using Eq. (\ref{velocities1}),
but 
we defer this manipulation to a later stage.

Let us also rewrite Eq.\ (\ref{density_def}) in terms of $\alpha$
and $F$:
\begin{equation}
\tilx_{n+1}-\tilx_n=\frac{1}{\alpha F\left[ (\tilx_{n+1}+\tilx_n)/2
\right]}\;.
\label{separations}
\end{equation}
According to this, the difference between successive $\tilx_n$'s is  
of order $\alpha^{-1}$ wherever $F$ does not vanish. This allows us to
expand Eq.\ (\ref{density_evolution2}) in powers of $\alpha^{-1}$ when
$\alpha$ is large and the step density is high. We can consider only
the leading order in $\alpha^{-1}$, 
an approximation which becomes exact in the scaling limit $\alpha
\rightarrow \infty$.  
The differences
$\tilx_{n+k}-\tilx$, where $\tilx$ denotes the middle of the terrace,
are also small
 as long as k is finite.
This allows us to go to a continuum limit in the variable $\tilx$ in the
following way. We evaluate the function $F$ at the position
$(\tilx_{n+k}+\tilx_{n+k+1})/2$ by using its Taylor expansion
\begin{equation}
  \label{taylor_expansion}
  F\left( \frac{ \tilx_{n+k}+\tilx_{n+k+1} } {2} \right)\equiv
  \frac{\alpha^{-1}}{\tilx_{n+k+1}-\tilx_{n+k}}=
  \sum_{m=0}^{\infty}\frac{1}{m!}\frac{d^m
  F(\tilx)}{d
  \tilx^m}\left(\frac{\tilx_{n+k}+\tilx_{n+k+1}}{2}-\tilx\right)^m\;.
\end{equation}
Now we expand
\begin{equation}
  \label{x_expansion}
  \tilx_{n+k}=\tilx+\sum_{m=1}^{\infty}\phi_{km}\alpha^{-m}\;,
\end{equation}
and insert the expansion into Eq.\ (\ref{taylor_expansion}). By
requiring all orders of 
$\alpha^{-1}$ to
cancel, the coefficients $\phi_{km}$ are found for any values of $k$ and
$m$. 
These coefficients involve the function $F$ and its
derivatives evaluated at $\tilx=(\tilx_n+\tilx_{n+1})/2$. 

We now return to Eq.\ (\ref{density_evolution2}). It depends on the
velocities $\dot{\tilx}_n$ and 
$\dot{\tilx}_{n+1}$, which in turn depend on
$\tilx_{n-2},\dots,\tilx_{n+3}$. Using Eq.\ 
(\ref{x_expansion}) we expand Eq.\ (\ref{density_evolution2}) in
$\alpha^{-1}$. The result 
of this expansion in the general NDL case is 
\begin{equation}
\frac{3}{4}\cdot \frac{d^2}{d\tilx^2}\left(\frac{1}{F}\cdot 
\frac{d^2F^2}{d\tilx^2}\right)+\frac{\dot{\alpha}}{\alpha} \cdot F+{\cal
O}\left(\alpha^{-2}\right)=0\;. 
\label{density_evolution3NDL}
\end{equation}
In the DL case the expansion result is different:
\begin{equation}
\frac{3}{2}\cdot \frac{d^4F^2}{d\tilx^4}+\frac{\dot{\alpha}}{\alpha^2}
\cdot F+{\cal O}\left(\alpha^{-2}\right)=0\;. 
\label{density_evolution3DL}
\end{equation}

Consider Eq.\ (\ref{density_evolution3NDL}) in the scaling limit,
$\alpha \rightarrow \infty$. 
The $\alpha^{-2}$ term can be ignored and  
we are left with the first two terms, which must cancel 
each other. This implies that $\dot{\alpha}/\alpha$ is independent
of time, i.e., in the NDL case 
\begin{equation}
\alpha \propto e^{-\tilt/\tau_{NDL}}\;,
\label{NDL_alpha}
\end{equation}
where $\tau_{NDL}$ is the life time of the profile. Note that the
scaling limit $\alpha \rightarrow \infty$ corresponds to very early
times $\tilt\rightarrow -\infty$.  
The equation for the NDL scaling function in this limit is  
\begin{equation}
\frac{3}{4}\cdot \frac{d^2}{d\tilx^2}\left(\frac{1}{F}\cdot 
\frac{d^2F^2}{d\tilx^2}\right)-\frac{F}{\tau_{NDL}}=0\;.  
\label{NDL_scaling_eq}  
\end{equation}
At any finite time when $\alpha$ itself is finite, there will be a
correction to scaling of order  
$\alpha^{-2}$. 

The same arguments can be applied to Eq.\ (\ref{density_evolution3DL})
in the DL case. We find that $\dot{\alpha}/\alpha^2$ is independent of
time, i.e.,
\begin{equation}
\alpha =\frac{\tau_{DL}}{\tilt-\tilt_0}\;.
\label{DL_alpha}
\end{equation}
In this case the scaling limit corresponds to a {\em finite} time,
$\tilt \rightarrow \tilt_0^+$. 
The differential equation for the DL case is
\begin{equation}
\frac{3}{2}\cdot \frac{d^4F^2}{d\tilx^4}-\frac{F}{\tau_{DL}}=0\;.
\label{DL_scaling_eq}
\end{equation}
Again there is a correction to scaling of order $\alpha^{-2}$ when
$t>t_0$.

Note that the scaling equation is the same for all NDL cases. Thus,
for all non-zero values of $l=\frac{2D_s}{k}$, the scaling behavior is
identical to that of the ADL limit. 
Since in the NDL case 
the terrace sizes vanish in the scaling limit, diffusion across terraces
is fast and,
unless the system if purely 
DL ($l=0$), attachment-detachment is 
always the rate limiting process. This  
fact makes the DL case special and somewhat delicate. Since in real
systems the diffusion coefficient is always finite, the scaling limit
of real systems will show ADL behavior. Nevertheless, at finite times we
can
expect crossover from ADL to DL behavior. This should happen at the
time when  
$\frac{1}{\alpha F} \approx l$. At later times, when $\frac{1}{\alpha F}
\gg l$, the
system should show DL behavior and obey Eq.\ (\ref{DL_scaling_eq}),
provided $\alpha$ is still large enough to neglect the $\alpha^{-2}$
corrections to scaling.

At this point we have a continuum model derived directly from the
discrete step equations of motion. In the 
scaling limit our model provides an exact connection between the
microscopic 
step kinetics and the macroscopic surface evolution. 
It is interesting 
to compare it with other continuum models, which do not
emerge from the underlying step model. 

In the usual continuum approach 
one assumes that surface dynamics is driven by the surface tendency to
decrease its {\em continuous} surface tension. Below the roughening
transition this surface tension has a cusp and takes the
form\cite{RettoriVillain,Nozieres}
\begin{equation}
\sigma(D)=\sigma_0+\sigma_1|D|+\sigma_3|D|^3\;,
\label{surface_tension}
\end{equation} 
with $D$ denoting the surface slope. The surface chemical potential is
consequently given by
\begin{equation}
\mu=\frac{\partial \sigma'(D)}{\partial x}\;,
\end{equation}
where $\sigma'$ is the derivative of $\sigma$ with respect to the
slope $D$.
This chemical potential gives rise to surface currents, which are
proportional to $\partial \mu/\partial x$ in the DL
case\cite{OzdemirZangwill,HagerSpohn,Nozieres}
and to $\partial \mu/ \partial h=\frac{1}{D}\frac{\partial
\mu}{\partial x}$ in the ADL case\cite{Nozieres}. The
equation of motion for the profile slope in regions where $D>0$ is
then given by 
\begin{equation}
\frac{\partial D}{\partial t} \propto
-3\sigma_3\frac{\partial^2}{\partial x^2}
\left(\frac{1}{D}\frac{\partial^2 D^2}{\partial x^2}\right)\;
\label{traditional_ADL}
\end{equation}
in the ADL case, and
\begin{equation}
\frac{\partial D}{\partial t} \propto -3\sigma_3\frac{\partial^4
D^2}{\partial x^4}\;
\label{traditional_DL}
\end{equation}
in the DL case.

To compare our model with the above equations, let us find the
general equation of 
motion for the density function; i.e., the partial
differential equation for the step density that by separation of
variables, $D(\tilx,\tilt)=\alpha(\tilt)F(\tilx)$, reduces to Eqs.\
(\ref{NDL_alpha})
\& (\ref{NDL_scaling_eq}) in the NDL case and to Eqs.\ (\ref{DL_alpha})
\& (\ref{DL_scaling_eq}) in the DL case. In the NDL case
we can replace 
$\tau_{NDL}$ in Eq.\ (\ref{NDL_scaling_eq}) by
$-\dot{\alpha}/\alpha$. The equation can then be written as
\begin{equation}
\frac{3}{4}\cdot
\frac{\partial^2}{\partial \tilx^2}\left(\frac{1}{D}\cdot 
\frac{\partial^2D^2}{\partial \tilx^2}\right)+\frac{\partial D}{\partial
\tilt}=0\;, 
\label{NDL_general_eq}  
\end{equation}
which is consistent with Eq.\ (\ref{traditional_ADL}) and confirms
Nozieres'
suggestion \cite{Nozieres} for ADL kinetics. 
Note that Eq.\ (\ref{NDL_general_eq}) applies only to ADL 
cases and not to the full range of NDL systems, because in the scaling
limit all NDL 
systems obey Eq.\ (\ref{NDL_scaling_eq}), which describes ADL
kinetics. Since Eq.\ (\ref{NDL_general_eq}) was derived from
 Eq.\ (\ref{NDL_scaling_eq}) it corresponds to ADL kinetics. 

For the DL case we replace $\tau_{DL}$ in Eq.\ (\ref{DL_scaling_eq}) by
$-\dot{\alpha}/\alpha^2$. We then have 
\begin{equation}
\frac{3}{2}\cdot
\frac{\partial^4 D^2}{\partial \tilx^4}+\frac{\partial D}{\partial
\tilt}=0\;,
\label{DL_general_eq}
\end{equation}
consistently with Eq.\ (\ref{traditional_DL}) and  Refs.\
\onlinecite{OzdemirZangwill,HagerSpohn} and \onlinecite{Nozieres}.  
Let us however remark that the validity of the
general Eqs.\
(\ref{NDL_general_eq}) and(\ref{DL_general_eq}) is questionable. In a
non scaling case these equations have corrections which  
may be important. We showed that in the scaling state these
corrections vanish and neglected them in moving from Eqs.\
(\ref{density_evolution3NDL}) \&  
(\ref{density_evolution3DL}) to (\ref{NDL_scaling_eq}) \&
(\ref{DL_scaling_eq}). In a general scenario, neglecting these terms is 
an approximation that must be justified.

\section{Solution of the continuum models}
In this section we find the scaling functions in the NDL and DL cases. 
Since the slope-up and slope-down sections are
symmetric, we need to consider only half a period, i.e., the section
$\tilx \in [-\pi/2,\pi/2]$. 
The scaling function is thus symmetric about the origin. 

We now discuss the possibility of macroscopic facets at the profile 
extrema. In terms of the discrete step model every terrace is a
facet. However, in the scaling limit the step density diverges with
$\alpha(\tilt)$, and the terraces in all regions where $F$ is finite
become 
truly microscopic. In regions where the
scaling function is identically zero the terrace size does not
necessarily vanish 
and macroscopic facets may form. To account for macroscopic facets we
should, in
principle, allow for regions of arbitrary size with $F=0$. However,
simulations of 
the discrete step model indicate that if such regions appear at all, 
they form around
the profile extrema where the step density is zero by definition.    
To account for macroscopic facets, we therefore allow for the existence
of two special points at $\pm \tilx_{facet}$, which denote the positions
of facet edges. In the regions $(-\pi/2,-\tilx_{facet})$ and
$(\tilx_{facet},\pi/2)$ the scaling function $F$ vanishes identically.
In this notation we can also describe systems without facets simply by 
setting $\tilx_{facet}=\pi/2$. Note that the scaling function should
satisfy Eq.\ (\ref{NDL_scaling_eq}) or (\ref{DL_scaling_eq}) only in
the interval $(-\tilx_{facet},\tilx_{facet})$ and not on the facet,
since
these equations are not valid when $F=0$. 

Next we show that if there is a macroscopic facet, $F$ is continuous
across the facet edge. To see this, consider
the annihilation process of the first step. 
Its velocity satisfies
\begin{equation}
\dot{\tilx}_1 \propto
\frac{(\tilx_3-\tilx_2)^{-3}-2(\tilx_2-\tilx_1)^{-3}-g(\tilx_1-\tilx_0)^
{-3}}{\frac{2 
\pi l}{\lambda}+\tilx_2-\tilx_1}\;.
\label{x1dot}
\end{equation}
The denominator in the above expression is always positive. Therefore
the
direction of motion of the first step is given by the sign of the
numerator,
which must be negative because the first step is moving towards 
annihilation with the zeroth step. Assume now that in the scaling
limit we have a macroscopic facet. This means that the distance
$\tilx_1-\tilx_0$ is finite when the annihilation process starts. Assume
also
that $F_{facet}=F\left(\tilx_{facet}\right)$ is finite. This makes the 
distance $\tilx_3-\tilx_2$  microscopically small, since in the
scaling limit it vanishes as $1/\left(\alpha F_{facet}\right)$. 
We can therefore neglect the last term in the numerator of Eq.\
(\ref{x1dot}). It is easy to see now that if
$\tilx_2-\tilx_1<2^{1/3}/ \left(\alpha F_{facet}\right)$ the velocity of
the first step is positive. Thus if $F_{facet}$ is finite, the first
step is
bounded to the second one at a distance which is much less than the
facet size. In such a situation the first step cannot annihilate. We
conclude that having a macroscopic facet ($\tilx_{facet}<\pi/2$)
requires the scaling 
function to vanish at the facet edge. Thus the scaling
function is continuous at $\tilx_{facet}$ with $F(\tilx_{facet})=0$.
Furthermore,
by expanding the 
scaling function in the vicinity of $\tilx_{facet}$, it can be shown
that 
both the NDL and DL scaling functions can be expressed as power series
of $\sqrt{\tilx_{facet}-\tilx}$:
\begin{equation}
F\left(\tilx\right)=\sum_{n=1}^{\infty}a_n\left(\sqrt{\tilx_{facet}-
\tilx}\right)^n\;.
\label{Fexpansion}
\end{equation}
Hence with a non-vanishing coefficient $a_1$, the derivatives of the
scaling function diverge at $\tilx_{facet}$.

Next we specify boundary conditions for the scaling function $F$.
Since Eqs.\ (\ref{NDL_scaling_eq}) and (\ref{DL_scaling_eq}) are
fourth order differential equations we need four boundary conditions in
order to solve them.
Three conditions are set by the normalization and symmetry of the
scaling function:
\begin{eqnarray}
&F&(0)=1\;, \nonumber \\
&F&'(0)=0\;, \nonumber \\
&F&'''(0)=0\;.
\label{bc123}
\end{eqnarray}

A fourth boundary condition can be found by considering
mass transfer in the decaying profile. Mass is
transfered from the decaying peaks to the valleys through the
sloping parts of the profile. The step-antistep symmetry of the system
excludes mass transfer through the profile extrema. We can thus
calculate the (dimensionless) flux through the origin by calculating
the (dimensionless) volume
change in the interval $[0,\pi/2]$:   
\begin{equation}
J(0)=\frac{d}{d\tilt}\int_0^{\pi/2}h(\tilx,\tilt)d\tilx\;.
\label{origin_flux1}
\end{equation}
$h(\tilx,\tilt)$ is the surface profile measured in steps. In the
continuum limit it is given by
\begin{equation}
h(\tilx,\tilt)=\int_0^{\tilx} D(\xi,\tilt) d\xi \;.
\label{cont_hight}
\end{equation}
Integrating Eq.\ (\ref{origin_flux1}) by parts and using the scaling
form 
(\ref{scaling_ansatz}) together with the possible existence of zero 
density facets results in
\begin{equation}
J(0)=\dot{\alpha}\int_0^{\tilx_{facet}}F
\cdot\left(\pi/2-\tilx\right)d\tilx\;.
\label{origin_flux2}
\end{equation}

To evaluate the integral in Eq.\ (\ref{origin_flux2}) in the NDL 
case, we multiply Eq.\ (\ref{NDL_scaling_eq}) by $(\pi/2-\tilx)$ and 
integrate from zero to $\tilx_{facet}$: 
\begin{equation}
\int_0^{\tilx_{facet}}F\cdot\left(\pi/2-\tilx\right)d\tilx=\frac{3
\tau_{NDL}}{4} \cdot \int_0^{\tilx_{facet}} 
\left(\pi/2-\tilx\right) \frac{d^2}{d\tilx^2} \cdot \left( \frac{1}{F}
\cdot \frac{d^2 F^2}{d\tilx^2} 
\right)d\tilx\;.
\end{equation}
The r.h.s.\ of the last equation can be integrated by parts. Inserting 
the result into Eq.\ (\ref{origin_flux2}) we find that
\begin{equation}
J(0)= \frac{3 \tau_{NDL} \dot{\alpha}}{4}\left.\left[ 
\left(\pi/2-\tilx\right)\frac{d}{d\tilx}\left(\frac{1}{F} \cdot
\frac{d^2 
F^2}{d\tilx^2}\right)+\frac{1}{F}\cdot \frac{d^2 F^2}{d\tilx^2}\right]
\right|_0^{\tilx_{facet}}\;.
\label{origin_flux3}
\end{equation}

Another method for calculating the flux through the origin relies on the 
discrete step system. The adatom flux across the $n$th terrace is
given by Eq.\  
(\ref{current}), which can be expanded in powers of $\alpha^{-1}$
using Eq.\ (\ref{x_expansion}). In our dimensionless variables, the
leading order of this expansion in the NDL case is 
\begin{equation}
J(\tilx)=\frac{3\alpha}{4F} \cdot \frac{d^2F^2}{d\tilx^2}\;.
\end{equation}
Evaluating this expression at $\tilx=0$ and comparing with Eq.\
(\ref{origin_flux3}), 
we obtain the boundary condition  
\begin{equation}
\left.\left[ \left(\pi/2-\tilx\right)\cdot
\frac{d}{d\tilx}\left(\frac{1}{F} \cdot 
\frac{d^2F^2}{d\tilx^2}\right) +\frac{1}{F} \cdot
\frac{d^2F^2}{d\tilx^2} \right] 
\right|_{\tilx_{facet}}=0\;.
\label{NDL_bc4}
\end{equation}
In deriving this boundary condition we used the fact that
$F'(0)=F'''(0)=0$ and  
that in the NDL case $\tau_{NDL}=-\alpha/\dot{\alpha}$.

The DL case is very similar. We can evaluate the  
flux through the origin both from Eq.\ (\ref{origin_flux2}) and from a 
direct 
expansion of Eq.\ (\ref{current}). Equating the two results we obtain
the fourth boundary condition in the DL case 
\begin{equation}
 \left.\left[ \left(\pi/2-\tilx\right)\cdot \frac{d^3F^2}{d\tilx^3} 
 +\frac{d^2F^2}{d\tilx^2} \right] \right|_{\tilx_{facet}}=0\;.
\label{DL_bc4}
\end{equation}

At this point we have four boundary conditions for the scaling
function, three at the origin and one at $\tilx_{facet}$. In addition we
know that  $F(\tilx_{facet})=0$ if $\tilx_{facet}<\pi/2$. We may now
find
unique scaling solutions if we know the values of $\tau_{NDL}$ and
$\tau_{DL}$, which enter in the differential equations
(\ref{NDL_scaling_eq}) and (\ref{DL_scaling_eq}). What determines the
values
of $\tau_{NDL}$ and $\tau_{DL}$? We now show that these life times
are related to the behavior of the top and bottom 
steps, which are unique in the sense that they each have a neighboring
step
of opposite sign. Our continuum model, which treats all steps on equal 
footing, does not contain any information about this unique behavior
and therefore $\tau_{NDL}$ and $\tau_{DL}$ must be calculated or
measured directly from the discrete step system.

To relate the profile life times to the behavior of the top and bottom 
steps, consider the number of steps in a half slope up region of the
profile. One can show from Eq. (\ref{cont_hight}) that it is given by  
\begin{equation}
N(t)=\int_0^{\pi/2}D(\tilx,\tilt)d\tilx=\alpha\int_0^{\tilx_{facet}}Fd
\tilx\;.
\label{step_number}
\end{equation}
The integral ${\cal I}=\int_0^{\tilx_{facet}}Fd\tilx$ can be evaluated
by integrating
Eqs. (\ref{NDL_scaling_eq}) or (\ref{DL_scaling_eq}) in the NDL or
the DL cases, respectively. The results are
\begin{eqnarray}
{\cal
I}_{NDL}&=&\left.\frac{3\tau_{NDL}}{4}\frac{d}{d\tilx}\left(\frac{1}{F}
\cdot 
\frac{d^2F^2}{d\tilx^2}\right)\right|_{\tilx_{facet}}\;, \nonumber \\
{\cal
I}_{DL}&=&\left.\frac{3\tau_{DL}}{2}\frac{d^3F^2}{d\tilx^3}\right|_
{\tilx_{facet}}\;.
\label{I_integral}
\end{eqnarray}
Combining Eqs. (\ref{step_number}) and (\ref{I_integral}) we can
calculate $\Delta \tilt$, the time interval in which the system looses
one
step, i.e., the annihilation time of the top step:
\begin{equation}
\Delta \tilt=\frac{d\tilt}{dN}=\frac{1}{\dot{\alpha}{\cal I}}\;.
\label{delta_t}
\end{equation}

Eq. (\ref{delta_t}) relates the profile decay rate, $\alpha$, to the
annihilation time of a single step at the profile peak, $\Delta \tilt$.
We could have
used this relation to set an additional condition at $\tilx_{facet}$ if
we 
knew $\Delta \tilt$. Such a condition would select a single solution
with
a specific life time from
the one dimensional family of possible solutions.
However, as we mentioned above, $\Delta
\tilt$ must be obtained directly from the discrete system, because our
continuum model ignores the unique behavior of the top step.

\begin{figure}[h]
\centerline{
\epsfxsize=80mm
\epsffile{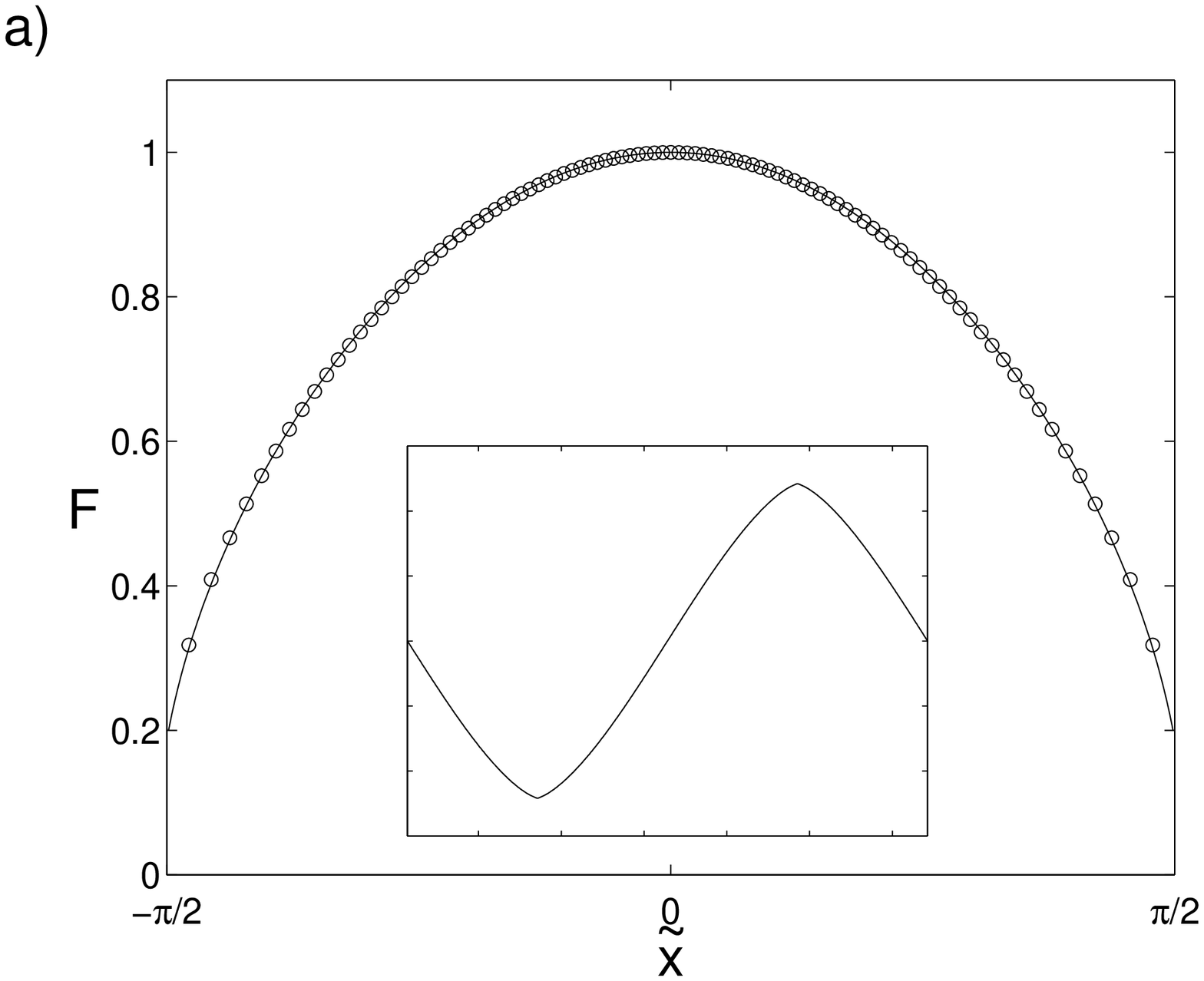}
\hspace{0.1in}
\epsfxsize=80mm
\epsffile{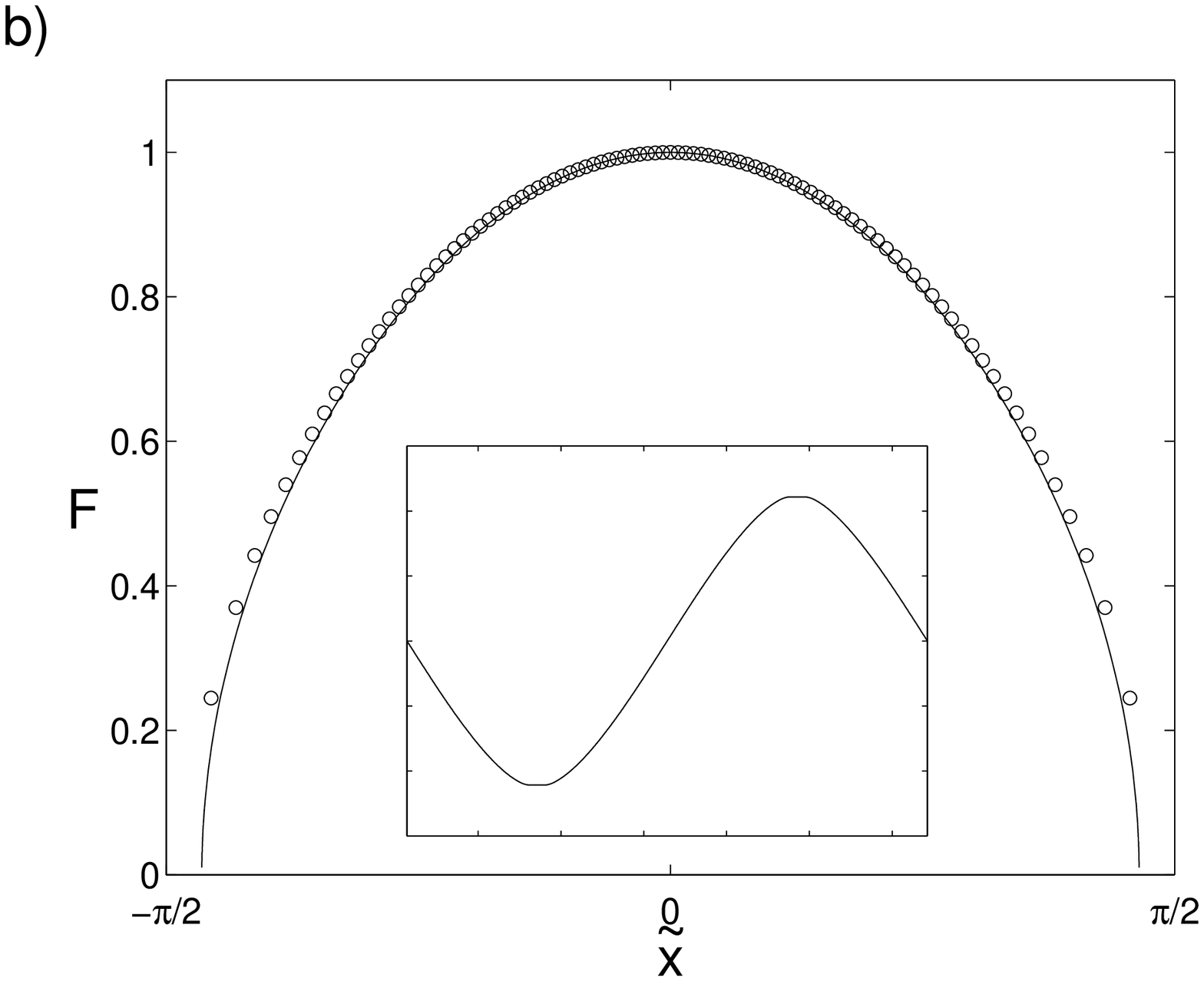}}
\vspace{0.1in}  
\caption{Scaling function solutions (solid) compared with simulation
data (circles) in the ADL case. Only a single slope up region is shown.
The insets
show the resulting surface profiles obtained by integrating the scaling 
functions over
alternating regions of steps and antisteps with a) $g=0$
and b) $g=24$.} 
\label{figure5}
\end{figure}

Unable to calculate the life times of the profiles analytically, we
measured $\tau_{NDL}$ and $\tau_{DL}$ from our simulations. Using
these values we solved Eqs.\ (\ref{NDL_scaling_eq}) and
(\ref{DL_scaling_eq}) numerically. In Fig.\ \ref{figure5} we show the
results for a single 
slope-up region of the profile in the ADL case and compare
them with the simulation data. The slope-down regions of the profile
behave in exactly the same way, but with a negative step density (to be
consistent with the profile slope). In the insets we 
show the surface profile, which is obtained by integrating the scaling 
function over
alternating regions of steps and antisteps. Similar results were
obtained in the DL case. 

The following picture emerges. Both in the ADL and DL cases, when
step-antistep attraction is absent 
($g=0$) the unique scaling solutions, which satisfy all the boundary
conditions, have no facets. The scaling functions obey Eqs.\
(\ref{NDL_scaling_eq}) and
(\ref{DL_scaling_eq}) in the entire interval $(-\pi/2,\pi/2)$ and are
finite at the
profile extrema ($\tilx=\pm \pi/2$). This
results in cusp-like peaks and valleys in the surface profile. The
agreement between the scaling functions and the simulation data is
excellent in these cases. When 
step-antistep attraction is present ($g=24$) the situation is different.  
The unique scaling solutions, which satisfy all the boundary conditions,
have small facets near the profile extrema. In the 
facet region the scaling functions are identically zero and as a result
the profiles have flat peaks and valleys. Here we also have good
agreement with simulation data, but not as good as in the $g=0$ cases.
This is because the scaling ansatz is a better approximation when
$g=0$.

\section{Connection with experiment}
In this section we test our model against
experimental results. Specifically we compare our
predictions with AFM measurements carried out by 
Tanaka et al\cite{BlakelyUmbachTanaka_snapshoot}. 
They measured the amplitude of decaying 1D Silicon gratings
fabricated on a 
surface vicinal to the Si(001) plane. They report an exponential
decay of the surface 
amplitude with life times scaling as the fourth power of the grating
wavelength. The observed profiles had flat peaks and valleys. These
results qualitatively resemble the behavior of our model in the NDL
case in the presence of step-antistep attraction.

To prepare the ground for a more quantitative comparison, let us first
discuss the relevance of our model to the experimental system. The
experimental conditions where such that the Si surface was below
its roughening transition. It is therefore reasonable to expect
step-dominated 
surface kinetics. However, the step structure in the experiment is
different from the idealized one dimensional array of straight
steps we considered in our model. Since the grating was fabricated on a
vicinal surface, with the grooves direction perpendicular to the
unperturbed steps, the steps are not straight. This can be easily shown
by writing the surface profile as
\begin{equation}
Z(x,y)=h(x)+ay\;.
\end{equation}
Here $h(x)$ is the one dimensional periodic profile along the $x$ axis
and $Z(x,y)$ is the two dimensional surface profile. The parameter $a$
measures the 
slope of the original vicinal surface with respect to the high symmetry
$(x,y)$ plane.   
The constant height contours which define the steps are given by 
\begin{equation}
y_n(x)=\frac{Z_n-h(x)}{a}\;.
\label{experimental_steps}
\end{equation}
Evidently, we deal here with curved steps, a fact which complicates 
surface dynamics. This situation was studied by Bonzel and Mullins
\cite{BonzelMullins} who investigated the smoothing of perturbed
vicinal surfaces. In addition to step-step
interactions, step 
motion is also driven by the steps line tension, $\Gamma$. 
This is reflected by the addition of a term
$\mu_n^{curv}(x)=\Gamma / R_n(x)$ to the step chemical potential,
where $R_n(x)$ is 
the local radius of curvature of the $n$th step.

The above arguments suggest that our straight steps model cannot
describe surface dynamics in regions where line tension is
important. To justify the application of our model to the experimental 
system we must therefore show that we can neglect the effect of step
line tension. We 
do this by considering the scaling limit of the system. We show
that in this limit the steps in the sloping parts of the
profile are effectively straight. The only regions where step
curvature is significant are the profile extrema, where our continuum
model is not valid anyway. Unlike Bonzel and Mullins we 
consider situations where step curvature is a driving force for
step straightening only at the profile extrema.

Assume now that the step system (\ref{experimental_steps}) exhibits
scaling and obeys Eq.\ (\ref{scaling_ansatz}). $D(x,t)=dh/dx$ is the
step
density along the $x$ axis. Under these assumptions we can express the
step curvature contribution to the chemical potential as 
\begin{equation}
\mu_n^{curv}(x)=\frac{\Gamma
y_n''}{\left[1+\left(y_n'\right)^2\right]^{3/2}}
=-\frac{\Gamma \alpha F'}{a \left(1+\frac{\alpha^2
F^2}{a^2}\right)^{3/2}}\;,
\label{curvature}
\end{equation}
with primes denoting derivatives with respect to $x$. Note that 
according to Eq.\ (\ref{curvature}), $\mu_n^{curv}(x) \equiv
\mu^{curv}(x)$ is independent of $n$.

Let us estimate the effect of $\mu^{curv}(x)$ on step
kinetics. This chemical potential gives rise to surface currents
parallel and perpendicular to the steps. The divergence of these
currents contributes to the step velocities. Along a step we
can estimate the resulting current as
$J_{\|} \propto d\mu^{curv}(x)/ds$ where $s$ is the step arclength.
The contribution of this current to the step velocity is
proportional to
\begin{eqnarray}
&&\frac{d^2 \mu^{curv}(x)}{ds^2}= \nonumber \\
&&\frac{\Gamma \,\left( -{\frac{\alpha \,F'''}{a}} + 
    {\frac{{{\alpha }^3}\,\left( 3\,{{F'}^3} + 10\,F\,F'\,F'' - 
          2\,{{F}^2}\,F''' \right) }{{a^3}}} + 
    {\frac{{{\alpha }^5}\,\left( -15\,{{F}^2}\,{{F'}^3} + 
          10\,{{F}^3}\,F'\,F'' - {{F}^4}\,F''' \right) }{{a^
         5}}} \right)}
{\left(1+\frac{\alpha^2 F^2}{a^2}\right)^{9/2}} \;.
\end{eqnarray}
In the scaling limit $\alpha \rightarrow \infty$, this
contribution decays as $\alpha^{-4}$ wherever $F$ is finite. This
result should be compared with the step velocities of the original
model. Using the expansion Eq.\ (\ref{x_expansion}), we can expand the
original step velocities, Eq.\ (\ref{velocities1}), in $\alpha^{-1}$. 
The outcome of this manipulation is a velocity of order $1$. Thus, in
the scaling limit, the curvature 
driven current along 
the steps results in a negligible contribution to the step velocities.  

The current perpendicular to a step is proportional to the chemical
potential
difference between two adjacent steps. This can be estimated from
$J_{\bot} \propto \hat{n} \cdot \nabla \mu^{curv}(x) / |\nabla Z|$
where $\hat{n}$ is a unit vector perpendicular to the step. However,
since we
are only interested here in the leading order in $\alpha^{-1}$, we can
make the approximation
$\hat{n}\approx x$, $|\nabla Z| \approx \alpha F$ and $J_{\bot}
\propto\frac{1}{\alpha F} \frac{d \mu^{curv}(x)}{dx}$ wherever $F$ does
not vanish. The
contribution of these currents to the step velocity is consequently
proportional to the current difference between two adjacent terraces
which we estimate as
\begin{equation}
\frac{1}{\alpha F}\frac{d}{dx}\left(\frac{1}{\alpha F}\frac{d
\mu^{curv}(x)}{dx}\right)=
-{\frac{\Gamma \alpha^{-4} a^2  \left( 15 F'^3 -
10FF'F'' + F^2F''' \right)
}{F^7}} 
+{\cal O}\left(\alpha^{-6}\right)\;.
\end{equation}
Again this contribution in negligible in the scaling limit.

\begin{figure}[h]
\centerline{
\epsfxsize=90mm
\epsffile{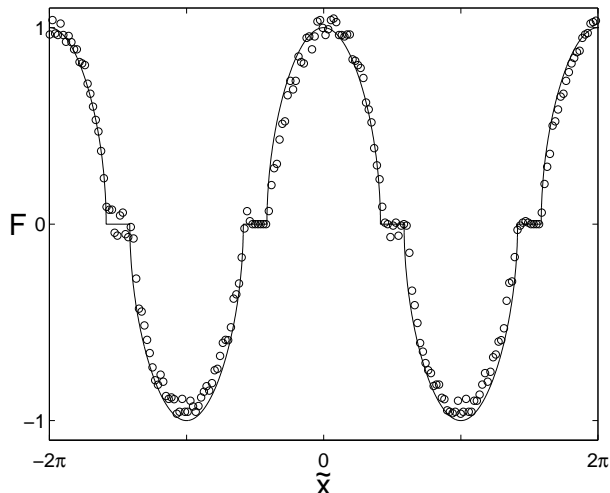}}  
\caption{Comparison between a snapshot of the experimental slope and
the best fit solution of Eq.\ (\protect{\ref{NDL_scaling_eq}}). This
solution
corresponds to a life time $\tau_{NDL}=0.945$. The experimental slope
was normalized to a $2\pi$ wavelength and unit maximal slope.}
\label{figure6}
\end{figure}

The above argument relies on the fact that in the sloping parts of the
profile $F$ is finite. This is of course not true near the profile
extrema. Step curvature is therefore important at the facets near the
profile peaks and valleys where the scaling function $F$ is
identically zero. Thus, we can divide the system into two types of
regions: the sloping parts where our continuum model is valid and the
facets where our continuum model breaks down. Recall however that our
model breaks down on facets in any case, and we solve the model only on   
the sloping parts of the profile. The kinetics of
the annihilating steps on the facets enters the continuum model
only through the life time of the profile, which we have to take
from the discrete step system. We can use this strategy here,
i.e., by measuring the life time of the profile from the experimental
results we can predict (using
Eq.\ (\ref{NDL_scaling_eq})) the density scaling
function. Alternatively, by fitting the scaling function to the
experiment we can make a connection between the experimental life time
$\tau_{exp}$, and
the microscopic parameters of the system. In Fig.\
\ref{figure6} we show a comparison between a snapshot of the
normalized experimental slope and the best fit solution of Eq.\
(\ref{NDL_scaling_eq}). This solution corresponds to a life time
$\tau_{NDL}=0.945$ in dimensionless units. According to Eq.\
(\ref{NDL_dimensionless_vars}) the three microscopic parameters
$k$, $\tilde{C}^{eq}$ and $\beta$ satisfy 
\begin{equation}
k\tilde{C}^{eq}\beta=\left(\frac{\lambda}{2\pi}\right)^4\frac{k_B T
\tau_{NDL}}{\Omega\tau_{exp}}\;.
\label{microscopic_parameters}
\end{equation}
The parameter values for the experimental data of Fig.\ \ref{figure6}
are $\lambda=5~\mu m$, $\tau_{exp}\approx 2.4\cdot 10^3$ minutes and
$T=900
^\circ C$.

Although Fig.\ \ref{figure6} shows good agreement
between theory and experiment it does not by itself provide enough
evidence for scaling in the experimental system. To verify that the
experimental system scales in time one must analyze several snapshots
of the surface slope taken at different times. However, the fit quality
and the fact that the surface amplitude decays exponentially with a
life time proportional to the fourth power of the grating wavelength,
strongly 
suggest scaling.

\section{Summary and Discussion}
In this work we have studied the relaxation process of one dimensional 
surface modulations below the roughening transition
temperature. Simulations of a step flow model suggest that the {\em
discrete} step density 
function, $D(x,t)$,  exhibits scaling throughout most of the decay
process, for a wide range of the model parameters,
spanning DL to ADL kinetics and a range of strengths of the attraction
between steps of opposite sign. 

Relying on the above observations, we then transformed the discrete step
model into a continuum model for surface dynamics. This was done using
the
scaling ansatz $D(x,t)=\alpha(t)F(x)$, and we were able to write down
the
differential equations for the functions $\alpha(t)$ and
$F(x)$. These equations were derived directly from the discrete step
equations of motion. We showed that they 
are exact in the scaling limit.

We investigated the
boundary conditions for the scaling function $F$ and found $F$
numerically.  A comparison between the
resulting solutions and density functions from simulations of the
discrete step model shows impressive agreement. 

Finally, we applied our continuum model to experimental data measured
by  Tanaka et al\cite{BlakelyUmbachTanaka_snapshoot}. They reported
exponential 
amplitude decay of Si(001)
gratings and profiles with flat peaks and valleys. In their experiment 
the profile life times scaled as the
fourth power of  
the grating wavelength. These properties resemble   
the behavior of our model in the NDL case with step-antistep attraction.
We showed that in the scaling limit, our one dimensional model is
adequate in spite of the fact that the steps 
in the experimental system are not straight. We compared
the best fit solution of our model to a normalized snapshot
of the experimental profile slope and found good agreement. On the
basis of this agreement  
we suggested that the profile exhibits scaling and made a connection
between the measured profile life time and microscopic system
parameters.

We now discuss the main aspects of our results and put them in
perspective with existing work. In the 
scaling limit our
continuum model is an exact result of step kinetics. We showed that the
resulting general differential equations ((\ref{NDL_general_eq}) and
(\ref{DL_general_eq})) are equivalent to existing phenomenological 
models\cite{OzdemirZangwill,HagerSpohn,Nozieres}, thereby
confirming that these models are consistent with step flow.  
Although we have not discussed this issue here, it can be shown that
the scaling solutions $D(x,t)=\alpha(t)F(x)$ are linearly stable
solutions of the general differential equations ((\ref{NDL_general_eq})
and
(\ref{DL_general_eq})). This fact may explain why scaling solutions
are selected.
However, our analysis shows that these
differential equations have corrections which originate from the
discrete nature of the system. When these corrections are important,
one must take into account the full expansion series
(\ref{x_expansion}). This results in differential equations of
infinite order. Therefore Eqs.\ (\ref{NDL_general_eq})
\& (\ref{DL_general_eq}) and other equivalent models are valid only
in situations where corrections can be ignored.  
We proved that the scaling scenario is such a case by showing how these
corrections vanish in the scaling limit.  

The second point we emphasize is that even in a scaling limit
there are regions where the discrete nature of steps becomes
important. Such regions are the profile extrema where step-antistep
pairs annihilate.  
The reason for this is twofold. First, the
steps near the profile extrema are unique in the sense that they have
a neighboring step of opposite sign. They therefore follow unique
equations of
motion. This does not enter in our continuum model which treats all
steps on equal footing. Secondly, our continuum model breaks down near 
zeros of the scaling function. This can be seen in the derivation of
the model, which relies on $F$ being finite. The kinetics of steps in
regions where 
the step density vanishes are not described by the model, since
corrections to the scaling function become important there. In
cases where the profile extrema are faceted, our continuum model
cannot account for the kinetics of steps on the facets even if
these steps follow the general discrete equations of motion.
 
It is therefore obvious that our continuum model (and all other
equivalent differential equations) does not constitute a complete
description of surface evolution. A manifestation of this statement is
the fact that the relative attraction strength, $g$, is not a parameter
of the continuum model, even though it strongly affects surface
evolution. Since our continuum model breaks down near the profile
extrema, their effect must be taken into account by supplying additional
information. In our case, this information is the value of the
profile life time.  As we showed in section IV, this life time
determines the
annihilation rate of step-antistep pairs and can be considered as the
last
boundary condition required for finding the scaling function. Thus, in a
scaling scenario, the effect of step kinetics near the profile
extrema reduces to a boundary condition for the scaling function at
$\pm \tilx_{facet}$. In order to solve for $F$ consistently with the
discrete model,
it is sufficient to supply the profile life time. We did this by
measuring the life times of the discrete systems.

In this respect our work is different from other continuum treatments of
modulated surfaces. As 
pointed out by Chame et al\cite{ChameRoussetBonzelVillain}, lack of
consistency with the kinetics  
of facet steps is the weakness of existing continuum models. One
remarkable outcome of these inconsistencies is the controversy
regarding the appearance of facets at the profile extrema.    
Ozdemir \& Zangwill\cite{OzdemirZangwill} and Hager \& Spohn
\cite{HagerSpohn} both used continuum models, which are equivalent to
Eq.\
(\ref{DL_general_eq}). Bonzel and Preuss \cite{BonzelPreuss} use a
slightly different model in which the surface currents are
proportional to the derivative of the chemical potential with respect
to the surface arclength. But they all start from a
continuum approach and do not specify the step kinetics on the facets.

Ozdemir and Zangwill assume that Eq.\
(\ref{DL_general_eq}) is valid in the full interval between the
profile extrema and thus exclude the possibility of facet formation. 
Their results show smooth
(but nonanalytic) profiles. This corresponds to some weak attractive
step-antistep interaction (in the absence of such interaction we get
cusped profiles). They showed that Eq.\ (\ref{DL_general_eq}) admits
shape preserving solutions, equivalent to our scaling
solutions in the DL case. In particular they showed that the amplitude
$\alpha$ of a shape preserving profile obeys Eq.\
(\ref{DL_alpha}). Our results for the decay rate in the DL case are
mathematically identical, but we interpret them differently. While
Ozdemir and Zangwill look at the long time limit in which $\alpha \sim
t^{-1}$, we recognize that Eq.\ (\ref{DL_general_eq}) is valid only
in the scaling limit, where $t \gtrsim t_0$. In this case $\alpha
\not\sim t^{-1}$. 

Hager and Spohn, on the other hand, allow for facet
formation by solving  Eq.\ (\ref{DL_general_eq}) with moving
boundaries. They assume that the chemical
potential, the current and the current divergence are all continuous
at the facet edge. In their model, the chemical potential at the facet 
edge is not a step chemical potential. It is a {\em layer} chemical
potential which reflects the free energy change caused by the
annihilation of
the top step-antistep pair. In this way they allow for the top
terraces to peel rapidly. Their model produces facets which grow in
time.

Our results are different. First, since we start from step
flow, the chemical potential in our continuum model is a {\em step}
chemical potential. On a facet our chemical potential
is not defined (and therefore not continuous) since there are no steps
there. One can argue that the adatom chemical potential on the
facet is equal to the chemical potential of the top first step. But
this first step is unique and is not treated by the continuum
model. It is also far from equilibrium with its neighbors so there
is no reason to assume continuity of the chemical potential.        
Second, the appearance of facets in our model depends on strength of
the step-antistep attraction.

Bonzel and Preuss deal with facets differently. They round the cusp
singularity of the surface tension, Eq.\ (\ref{surface_tension}), and
apply
the kinetic surface 
equation everywhere. They thus obtain analytic profiles with very flat
extrema
but without actual faceting.
Again the weakness of this approach is that it is not clear how this
rounding is related to the microscopic behavior of steps. 
 
We conclude that our continuum model is fully consistent with step
kinetics, but is restricted 
to scaling scenarios. It may serve both as a starting point and a test
case, for new continuum models which will attempt to describe surface
evolution in general and facet kinetics in particular.    

We are greatful to C. C. Umbach for supplying the experimental data and
for helpful discussions.
This research was supported by grant No. 95-00268 from the
United States-Israel Binational Science Foundation (BSF), Jerusalem,
Israel.


\begin{thebibliography}{99}


\bibitem[*]{NavotEmail}
E-mail: israeli@wicc.weizmann.ac.il

\bibitem[**]{DanielEmail}
E-mail: daniel.kandel@weizmann.ac.il, 
http://www.weizmann.ac.il/\~{ }fekandel.



\bibitem{Yamashita}
K. Yamashita, H. P. Bonzel and H. Ibach, Appl. Phys. {\bf 25}, 231
(1981); 

\bibitem{Mullins}
W. W. Mullins, J. Appl. Phys. {\bf 28}, 333 (1957); {\bf 30}, 77 (1959).

\bibitem{BlakelyUmbachTanaka_snapshoot}
S. Tanaka, C. C. Umbach, J. M. Blakely, R. M. Tromp and M. Mankos, 
J. Vac. Sci. Technol. A {\bf 15(3)} 1345 (1997). 

\bibitem{Tanaka}
S. Tanaka, C. C. Umbach, J. M. Blakely, R. M. Tromp and M. Mankos, 
Mat. Res. Soc. Symp. Proc. {\bf 440} 25 (1997).

\bibitem{Blakely}
J. Blakely, C. Umbach and S. Tanaka, Dynamics of Crystal Surfaces and
Interfaces, edited by P. M. Duxbury and T. J. Pence (Plenum, New York,
1997), p. 23.  

\bibitem{Umbach}
C. C. Umbach, M. E. Keeffe and J. M. Blakely, J. Vac. Sci. Technol. A
{\bf 9}, 1014 (1991).


\bibitem{OzdemirZangwill}
M. Ozdemir and A. Zangwill, Phys. Rev. B {\bf 42}, 5013 (1990).

\bibitem{RettoriVillain}
A. Rettori and J. Villain, J. Phys. France {\bf 49}, 257 (1988).

\bibitem{DuportChameMullinsVillain}
C. Duport, A. Chame, W. W. Mullins and J. Villain, J. Phys. I France
{\bf 6}, 1095 (1996).



\bibitem{RamanaCooper}
M. V. Ramana Murty and B. H. Cooper, Phys. Rev. B {\bf 54}, 10377
(1996). These Monte Carlo simulations agree
with{\protect{\cite{RettoriVillain,OzdemirZangwill}}}. 

\bibitem{AdamChameLanconVillain}
E. Adam, A. Chame, F. Lancon and J. Villain, J. Phys. I France
{\bf 7}, 1455 (1997).

\bibitem{DubsonJeffers}
M. A. Dubson and G. Jeffers, Phys. Rev. B {\bf 49}, 8347 (1994).
 
\bibitem{JiangEbner}
Z. Jiang and C. Ebner, Phys. Rev. B {\bf 40}, 316 (1989).

\bibitem{SelkeDuxbury}
W. Selke and P. M. Duxbury, Phys. Rev. B {\bf 52}, 17468 (1995).

\bibitem{ErlebacherAziz}
J. D. Erlebacher and M. J. Aziz, Surf. Sci. {\bf 374}, 427
(1997). 



\bibitem{HagerSpohn}
J. Hager and H. Spohn, Surf. Sci. {\bf 324}, 365 (1995).

\bibitem{BonzelPreussSteffen}
H. P. Bonzel, E. Preuss and B steffen, Appl. Phys. A {\bf 35}, 1 (1984).

\bibitem{BonzelPreuss}
H. P. Bonzel and E. Preuss, Surf. Sci. {\bf 336}, 209 (1995).

\bibitem{ChameRoussetBonzelVillain}
A. Chame, S. Rousset, H. P. Bonzel and J. Villain, Bul. Chem. Comm. {\bf
29}, 398 (1996/1997).



\bibitem{cone_short}
N. Israeli and D. Kandel, Phys. Rev. Lett. {\bf 80} 3300 (1998).

\bibitem{cone_long}
N. Israeli and D. Kandel, Phys. Rev. B {\bf 60} 5946 (1999).

\bibitem{1D_scaling}
N. Israeli, H-C. Jeong, D. Kandel and J. D. Weeks, Phys. Rev. B {\bf 61}
5698 (2000).  




\bibitem{BCF}
W. K. Burton, N. Cabrera and F. C. Frank,
Philos. Trans. R. Soc. London, Ser. A {\bf 243}, 299 (1951).


\bibitem{BalesZangwill}
G. S. Bales and A. Zangwill, Phys. Rev. B {\bf 41} 5500 (1990).




\bibitem{MarchenkoParshin}
V. I. Marchenko and A. Ya. Parshin, Zh. Eksp. Teor. Fiz. {\bf 79}, 257
(1980) [Sov. Phys. JETP {\bf 52}, 129 (1980)].

\bibitem{AndreevKosevich}
A. F. Andreev and Yu. A. Kosevich, Zh. Eksp. Teor. Fiz. {\bf 82}, 1435
(1981) [Sov. Phys. JETP {\bf 54}, 761 (1982)].


\bibitem{Nozieres}
P. Nozieres, J. Phys. I France {\bf 48}, 1605 (1987).


\bibitem{BonzelMullins}
H. P. Bonzel and W. W. Mullins, Surf. Sci. {\bf 350}, 285 (1996). 




\bibitem{LanconVillain}
F. Lancon and J. Villain, Phys. Rev. Lett. {\bf 64}, 293 (1990).







  

\end{thebibliography}
\end{document}